\documentclass[aps,prb,twocolumn]{revtex4} 

\usepackage{graphicx}
\usepackage{dcolumn}
\usepackage{bm}

\newcommand{\comment}[1]{}

\begin{document}

\title{Quantum Statistical Mechanics in Classical Phase Space.
III.
Mean Field Approximation Benchmarked
for Interacting Lennard-Jones Particles}


\author{Phil Attard}
\affiliation{phil.attard1@gmail.com Sydney NSW, Australia}


\begin{abstract}
A Monte Carlo computer simulation algorithm in classical phase space
is given for the treatment of quantum systems.
The non-commutativity of position and momentum
is accounted for by a mean field approach
and instantaneous effective harmonic oscillators.
Wave function symmetrization is included
at the dimer and double dimer level.
Quantitative tests are performed
against benchmarks given by Hernando and Van\'i\v cek (2013)
for  spinless neon--parahydrogen,
modeled as interacting Lennard-Jones particles
in a one dimensional harmonic trap.
The mean field approach is shown to be quantitatively accurate
for high to moderate temperatures
$\beta \hbar \omega_\mathrm{LJ}  \alt 7$,
and moderate densities, $\rho \sigma \approx 1$.
Results for helium
show that at the lowest temperature studied,
the average energy is about 4\% lower for bosons than for fermions.
It is argued that the mean field
algorithm will perform better in three dimensions than in one,
and that it will scale sub-linearly with system size.
\end{abstract}


\maketitle

%
\section{Introduction}
\setcounter{equation}{0} \setcounter{subsubsection}{0}
\renewcommand{\theequation}{\arabic{section}.\arabic{equation}}
%

The challenges posed by quantum condensed matter systems
to computational description
are well documented.
\cite{Parr94,Morton05,Bloch08,McMahon12,Pollet12,Hernando13}
These will not be further reviewed here,
except to note generically that
the problems arise primarily
from the difficulty in finding the energy
eigenfunctions and eigenvalues,
and from the hardship of the boson and fermion occupancy rules.
These impose technical barriers
to developing accurate numerical algorithms,
and create practical impediments such as
the rapid increase in computational cost with system size.
Usually approximations have to be introduced
such as neglecting wave function symmetrization,
using approximate eigenfunctions,
working on  a lattice,
or using a reduced set of eigenstates.
Such approximations,
while necessary and justifiable is specific cases,
can limit the reliability and range of application of any given method.

In view of the difficulties with existing approaches,
the present author has taken a different path.
Quantum statistical mechanics has been formulated
in classical phase space,\cite{QSM,STD2}
which offers advantageous scaling with system siz,
 considerably reduced computational demands,
and access to the many algorithms that have been developed
for classical systems.
The specific effects
that differentiate quantum from classical systems,
namely wave function symmetrization,
which gives rise to boson and fermion statistics,
and also the non-commutativity or lack of simultaneity
of position and momentum,
are accounted for with formally exact phase space expressions.
\cite{Attard18a}

The analysis invokes directly position and momentum states,
and is a somewhat simpler formulation
than the method of Wigner,\cite{Wigner32}
and of Kirkwood.\cite{Kirkwood33}
The author's approach is not directly related
to these earlier approaches,
although it agrees  with them for the first and second quantum
corrections to classical statistical mechanics.\cite{QSM,STD2}
The author's approach
has been tested analytically and numerically
for the quantum ideal gas,\cite{QSM,STD2}
and for non-interacting quantum simple harmonic oscillators.\cite{Attard18b}
In both cases there was agreement
with the known exact analytic results.
\cite{Messiah61,Merzbacher70,Pathria72}

The present paper applies the author's approach
to a non-ideal system of interacting Lennard-Jones particles.
The results are tested quantitatively against the benchmark results
given by Hernando and Van\'i\v cek,\cite{Hernando13}
who obtained the first fifty energy eigenfunctions and eigenvalues
of a system of 4 or 5 interacting spinless Lennard-Jones particles
trapped in a one-dimensional harmonic potential.
The parameters correspond to a hypothetical particle
with properties between parahydrogen and neon.\cite{Hernando13}
In addition to the tests against these benchmark results,
the present paper also reports results for the same system with helium.

The present algorithm consists of classical Metropolis Monte Carlo,
combined with umbrella sampling for the commutation function
and symmetrization function weight.
The commutation function is evaluated in mean field,
analytic, harmonic oscillator approximation,
as defined by the instantaneous local curvature
of the potential energy.\cite{Attard18b}
This mean field quantum harmonic oscillator approach
differs from an unrelated perturbation method
that estimates quantum corrections to a classical system
by using the spectrum of the classical velocity
auto-correlation function  as the density of states
for a distribution of effective quantum simple harmonic oscillators.
\cite{Berens83}

The major purpose of this paper is to establish the regime of validity
of the mean field approximation by comparison with the benchmark results.
The symmetrization function is evaluated in a loop expansion
that retains the monomer, dimer, and double dimer terms.
Results are presented for both fermions and bosons.
Although the differences between them are small for neon-parahydrogen
(but more noticeable for helium),
they nevertheless  justify the conclusion that
the symmetrization loop expansion is rapidly converging,
and that the additional computational burden
in accounting for particle statistics is not prohibitive
in the present phase space formulation of quantum statistical mechanics.

%
\section{Phase Space Formulation of Quantum Statistical Mechanics}
\label{Sec:qp}
\setcounter{equation}{0} \setcounter{subsubsection}{0}
%

\subsection{Exact Formulation}

The fundamental formulae upon which the simulation algorithm
is based is summarized here;
a full derivation is given in Ref.~\onlinecite{Attard18a}.
For a canonical equilibrium quantum system,
the statistical average in phase space is\cite{Attard18a}
\begin{eqnarray} \label{Eq:<A>}
\left< \hat A \right>_{N,V,T}^\pm
& = &
\frac{1}{Z^\pm}
\mbox{TR}' \left\{ e^{-\beta \hat{\cal H}} \hat A \right\}
\nonumber \\ & = &
\frac{1}{Z^\pm h^{dN} N!}
\int \mathrm{d}{\bf \Gamma}\;
\nonumber \\ && \mbox{ } \times
e^{-\beta{\cal H}({\bf \Gamma})} A({\bf \Gamma})
W_{p}({\bf \Gamma}) \eta^\pm_q({\bf \Gamma}) .
\end{eqnarray}
Here $\beta = 1/k_\mathrm{B}T$ is the inverse temperature of the reservoir,
with $k_\mathrm{B}$ being Boltzmann's constant and $T$ being the temperature,
$N$ is the number of  particles, assumed identical,
$V$ is the volume,
and $d$ is the dimensionality, all for the sub-system.
Also $h$ is Planck's constant, 
${\bf \Gamma} = \{ {\bf p},{\bf q}\}$ is a point in classical phase space,
with the position vector being
${\bf q} = \{{\bf q}_1, {\bf q}_2, \ldots , {\bf q}_N \}$,
and ${\bf q}_j = \{ q_{jx}, q_{jy}, \ldots, q_{jd} \}$,
and similarly for the momentum vector ${\bf p}$,
and
 ${\cal H}({\bf \Gamma}) = {\cal K}({\bf p}) + U({\bf q})$
 is the classical Hamiltonian or total energy function,
with ${\cal K}({\bf p}) = p^2/2m$ being the kinetic energy
($m$ is the mass of the particles),
and $U({\bf q})$ being the potential energy.
Finally, $Z^\pm$ is the normalizing canonical partition function,
with the superscript designating the type of particle:
the upper sign is for bosons and the lower sign is for fermions.

The commutation function $W$
arises from the non-commutativity of the position and momentum operators.
\cite{QSM,STD2}
It is essentially the same as the function
introduced by Wigner\cite{Wigner32}
and analyzed by Kirkwood,\cite{Kirkwood33}
and is defined by
\begin{eqnarray}\label{Eq:def-Wp}
e^{-\beta{\cal H}({\bf \Gamma})}
W_{p}({\bf \Gamma})
& = &
\frac{\langle{\bf q}|e^{-\beta\hat{\cal H}} |{\bf p}\rangle
}{\langle{\bf q}| {\bf p}\rangle }
\nonumber \\ & = &
\frac{1}{\langle{\bf q}| {\bf p}\rangle }
\sum_{\bf n} e^{-\beta E_{\bf n}}
\langle{\bf q} | {\bf n} \rangle \,
\langle{\bf n} | {\bf p} \rangle .
\end{eqnarray}
One has
$W_p({\bf p},{\bf q})^* =
W_p(-{\bf p},{\bf q})$,
where the asterisk denotes the complex conjugate.
High temperature expansions for the commutation function
have been given.\cite{Wigner32,Kirkwood33,STD2,Attard18a}
The second equality invokes
the sum over the energy eigenstates of the sub-system,
with the energy eigenfunctions in the position and momentum
representations appearing.
The present simulations are based on the commutation function in this form,
as will be discussed below.

In general the commutation function is specific
to the function being averaged.
But for functions that are linear combinations
of functions purely of momentum or purely of position,
this generic form is valid.\cite{Attard18a}
Energy and density, which are the two major quantities
sought in the simulations reported below, fall into this category.

The symmetrization function $\eta^\pm$
arises from the fact that
the wave function must be fully symmetric (bosons, upper sign)
or fully anti-symmetric (fermions, lower sign)
with respect to particle permutation.\cite{Messiah61,Merzbacher70}
Formally it is\cite{QSM,STD2,Attard18a}
\begin{eqnarray}
\eta^\pm_q({\bf \Gamma})
& \equiv &
\frac{1}{\langle {\bf p} | {\bf q} \rangle }
\sum_{\hat{\mathrm P}} (\pm 1)^p \,
\langle \hat{\mathrm P} {\bf p} | {\bf q} \rangle
\nonumber \\ & = &
e^{{\bf p}\cdot{\bf q}/i \hbar}
\sum_{\hat{\mathrm P}} (\pm 1)^p \,
e^{- (\hat{\mathrm P}{\bf p}) \cdot{\bf q}/i \hbar} .
\end{eqnarray}
Here $\hat{\mathrm P}$ is the permutation operator
and $p$ is its parity.
The second equality invokes the momentum eigenfunctions
in the position representation,
$|{\bf p}\rangle \equiv
{e^{-{\bf p}\cdot{\bf r}/i \hbar}}/{V^{N/2}}$,
with the position eigenfunctions being
Dirac delta functions,
$|{\bf q}\rangle \equiv \delta(({\bf r}-{\bf q})$.
Evidently,
$\eta^\pm_q({\bf p},{\bf q})^* =
\eta^\pm_q(-{\bf p},{\bf q})$.
Because of the exponential terms,
the symmetrization function is highly oscillatory,
which means it cancels to zero unless the particles involved
in any given permutation are close together in phase space.
This gives rise to a rapidly converging loop expansion,
\cite{QSM,STD2,Attard18a}
which is discussed and exploited below.

\subsection{Mean Field Harmonic Oscillator Approximation}

In the simulations reported below,
a mean field approximation is used
that exploits the analytic form
of the commutation function in the case of independent simple
harmonic oscillators.\cite{Attard18b}
This approximation is now described.

\subsubsection{Mean Field Approximation}

In general, the particles of the sub-system
interact via the potential energy,
which is the sum of one-body,
two-body, three-body terms, etc.,
\begin{eqnarray}
U({\bf q}) & = &
\sum_{j=1}^N u^{(1)}({\bf q}_j)
+ \sum_{j<k}^N u^{(2)}({\bf q}_j,{\bf q}_k)
\nonumber \\ && \mbox{ }
+ \sum_{j<k<l}^N u^{(3)}({\bf q}_j,{\bf q}_k,{\bf q}_l)
+ \ldots
\end{eqnarray}
Distributing the energy equally,
the energy of particle $j$ 
can be defined as
\begin{eqnarray}
U_j({\bf q}_j;{\bf q})
& = &
u^{(1)}({\bf q}_j)
+ \frac{1}{2} \sum_{k=1}^N\!^{(k\ne j)} \, u^{(2)}({\bf q}_j,{\bf q}_k)
\nonumber \\ && \mbox{ }
+ \frac{1}{3}
\sum_{k<l}^N\!^{(k,l\ne j)} \, u^{(3)}({\bf q}_j,{\bf q}_k,{\bf q}_l)
+ \ldots
\end{eqnarray}
The total potential energy is just
$ U({\bf q}) = \sum_{j=1}^N U_j({\bf q}_j;{\bf q})$.
The argument $({\bf q}_j;{\bf q})$ means that ${\bf q}_j$
is here separated out from ${\bf q}$.

The potential energy of particle $j$ in configuration ${\bf q}$
may be expanded to second order
about its local minimum at $\overline{\bf q}_j({\bf q})$,
\begin{equation}
U_j({\bf q}_j;{\bf q})
=
\overline U_j({\bf q})
+ \frac{1}{2} [{\bf q}_j-\overline{\bf q}_j][{\bf q}_j-\overline{\bf q}_j]
: \overline{\underline{\underline U}}_j'' ,
\end{equation}
where the minimum  value of the potential is
$\overline U_j({\bf q}) \equiv U_j(\overline{\bf q}_j;{\bf q})$.
The gradient,
$\nabla_j U_j({\bf q}_j;{\bf q})$,
vanishes at ${\bf q}_j = \overline{\bf q}_j({\bf q})$.
The  $d \times d$ second derivative matrix for particle $j$
at the minimum,
$\overline{\underline{\underline U}}_j''
= \left. \nabla_j \nabla_j  U_j({\bf q}_j;{\bf q})
 \right|_{{\bf q}_j = \overline{\bf q}_j}$,
is assumed positive definite.
Computationally,
it is convenient to iterate Newton's method
to locate the local minimum,
\begin{equation}
\overline{\bf q}_j^{(n+1)}
\approx
\overline{\bf q}_j^{(n)}
-\left[{\underline{\underline U}}_j''(\overline{\bf q}_j^{(n)})\right]^{-1}
\nabla_j U_j(\overline{\bf q}_j^{(n)})  .
\end{equation}

For configurations ${\bf q}$
that have no local minimum in the potential,
or that have too large a displacement $|{\bf q}_j- \overline{\bf q}_j|$,
the corresponding single particle commutation function
can be set to unity, $W_j({\bf \Gamma}) = 1$.
This is justified by analytic results
for the simple harmonic oscillator.\cite{Attard18b}

The positive definite second derivative matrix
has $d$ eigenvalues $\lambda_{j\alpha}({\bf q})  > 0$,
and  orthonormal eigenvectors,
$ \overline{\underline{\underline U}}_j'' {\bf X}_{j\alpha}
= \lambda_{j\alpha}{\bf X}_{j\alpha}$, $\alpha = x,y,\ldots, d$.
As $d$ is typically 1, 2, or 3, it is trivial numerically
to find the eigenvalues and eigenvectors and to diagonalize
$\overline{\underline{\underline U}}_j''$.
For molecule $j$ in configuration ${\bf q}$ the eigenvalues
define the frequencies
\begin{equation}
\omega_{j\alpha}({\bf q}) = \sqrt{ \lambda_{j\alpha}({\bf q}) /m }
, \;\; \alpha = x, y, \ldots , d .
\end{equation}
With this the potential energy is
\begin{eqnarray}
U({\bf q})
& = &
\sum_{j=1}^N \overline U_j
+
 \frac{1}{2}\sum_{j=1}^N
({\bf q}_j-\overline{\bf q}_j) ({\bf q}_j-\overline{\bf q}_j) :
\overline{\underline{\underline U}}_j''
\nonumber \\ & = &
\sum_{j=1}^N  \overline U_j
+
\frac{1}{2} \sum_{j,\alpha}  \hbar \omega_{j\alpha} Q_{j\alpha}^2 .
\end{eqnarray}
Here $ Q_{j\alpha} \equiv
\sqrt{ { m \omega_{j\alpha} }/{\hbar} } \, Q_{j\alpha}' $,
and
$ {\bf Q}_j' =
 \underline{\underline X}_j^\mathrm{T}  [{\bf q}_j-\overline{\bf q}_j] $.

\subsubsection{Harmonic Oscillator Analysis}

The mean field approximation combined with
the second order expansion about the local minima
maps each configuration ${\bf \Gamma}$
to a system of $dN$ independent harmonic oscillators
with frequencies $\omega_{j\alpha}$
displacements $Q_{j\alpha}$,
and momenta ${P}_{j\alpha}
= \sqrt{m\hbar \omega_{j\alpha}}
\left\{ \underline{\underline X}_j^\mathrm{T}
{\bf p}_j \right\}_\alpha$.

With this harmonic approximation
for the potential energy,
the effective Hamiltonian in a particular configuration can be written
\begin{equation}
{\cal H}^\mathrm{SHO}({\bf p},{\bf q}-\overline{\bf q})
=
\sum_{j=1}^N \overline U_j
+
\frac{1}{2} \sum_{j,\alpha}  \hbar \omega_{j\alpha}
\left[ P_{j\alpha}^2 + Q_{j\alpha}^2 \right] .
\end{equation}

The commutation function for the interacting system
for a particular configuration
can be approximated as the product of commutation functions
for effective non-interacting harmonic oscillators
which have the local displacement as their argument.
With this the mean field commutation function is
\begin{eqnarray}
W_p^\mathrm{mf}({\bf \Gamma})
& \approx &
W_p^\mathrm{SHO}({\bf p},{\bf q}-\overline{\bf q})
\nonumber \\ & = &
e^{\beta {\cal H}^\mathrm{SHO}({\bf p},{\bf q}-\overline{\bf q})}
\frac{ \langle {\bf q}-\overline{\bf q}
|e^{-\beta \hat{\cal H}^\mathrm{SHO} }
|{\bf p}\rangle
}{ \langle {\bf q}-\overline{\bf q}|{\bf p}\rangle }
\nonumber \\ & = &
\prod_{j,\alpha} W_{p,j\alpha}^\mathrm{SHO}(P_{j\alpha},Q_{j\alpha}) .
\end{eqnarray}
The harmonic oscillator commutation function
for a single mode is\cite{Attard18b}
\begin{eqnarray} \label{Eq:WSHO}
\lefteqn{
W_{p,{j\alpha}}^\mathrm{SHO}(P_{j\alpha},Q_{j\alpha})
}  \\
& = &
\sqrt{2}
e^{-iP_{j\alpha}Q_{j\alpha}}
e^{\beta\hbar \omega_{j\alpha} [P_{j\alpha}^2+Q_{j\alpha}^2]/2 }
e^{-[P_{j\alpha}^2+Q_{j\alpha}^2]/2 }
\nonumber \\ && \mbox{ } \times
\sum_{n_{j\alpha}=0}^\infty
\frac{i^{n_{j\alpha}}
e^{- \beta \hbar \omega_{j\alpha}  (n_{j\alpha}+1/2) }
}{
2^{n_{j\alpha}} {n_{j\alpha}}!  }
\mathrm{H}_{n_{j\alpha}}(P_{j\alpha})
\mathrm{H}_{n_{j\alpha}}(Q_{j\alpha}) .\nonumber
\end{eqnarray}
The prefactor $e^{-iP_{j\alpha}Q_{j\alpha}}$
corrects the prefactor $e^{-ip_{j\alpha}q_{j\alpha}/\hbar}$
given in  Eq.~(5.10) of Ref.~[\onlinecite{Attard18b}].
Here $\mathrm{H}_n(z)$ is the Hermite polynomial of degree $n$.
The imaginary terms here are odd in momentum.
As mentioned, for configurations
such that $\overline{\underline{\underline U}}_j''$
is not positive definite,
or that the displacement $Q_{j\alpha}$ is too large,
the commutation function can be set to unity,
$W_{p,{j\alpha}}^\mathrm{SHO} = 1$.

For the averages, the momentum integrals
can be performed analytically,
both here and in combination with the symmetrization function.
This considerably reduces computer time
and substantially increases accuracy.

\subsection{Symmetrization Function}

The symmetrization function can be written as a series of loop products,
\cite{STD2,QSM,Attard18a}
\begin{eqnarray}
\eta^\pm_q({\bf \Gamma})
& = &
1
+ \sum_{jk}\!'  \eta_{q;jk}^{\pm(2)}
+ \sum_{jkl}\!'  \eta_{q;jkl}^{\pm(3)}
\nonumber \\ &&  \mbox{ }
+ \sum_{jklm}\!' \eta_{q;jk}^{\pm(2)}
\eta_{q;lm}^{\pm(2)}
+ \ldots
\end{eqnarray}
Here the superscript is the order of the loop,
and the subscripts are the atoms involved in the loop.
The prime signifies that the sum is over unique loops
(ie.\ each configuration of particles in loops occurs once only)
with each index different
(ie.\ no particle may belong to more than one loop).
The $l$-loop symmetrization factor is
\begin{equation} \label{Eq:tilde-eta-l}
\eta_{q;1 \ldots l}^{\pm(l)}
=
(\pm 1)^{l-1}
e^{ {\bf q}_{l1} \cdot {\bf p}_1 /i\hbar }
\prod_{j=1}^{l-1}
e^{ {\bf q}_{j+1,j} \cdot {\bf p}_j /i\hbar } ,
\end{equation}
where $ {\bf q}_{jk} \equiv {\bf q}_{j}- {\bf q}_{k}$.

In the series loop expansion above,
the first term of unity is for monomers.
The second term is a dimer loop,
the third term is a trimer loop,
and the fourth term shown is the product of two dimers.
The monomer term, $\eta^{\pm(1)} = 1$,
is obviously the classical one,
and it is the only term present for so-called distinguishable particles.
This is the dominant term when $\rho \Lambda^d_\mathrm{th} \ll 1$,
which is the low density, high temperature limit.
Here $\rho=N/V$ is the number density
and $\Lambda_\mathrm{th} = [2\pi \hbar^2 \beta/m]^{1/2}$
is the de Broglie thermal wave length.

As mentioned, these loop symmetrization factors are highly oscillatory,
and so the only non-zero contribution from them comes
from configurations such that successive particles around a loop
are close neighbors in phase space.
In previous work\cite{STD2,QSM,Attard18a}
the compact nature of the loops
was used to argue that in the thermodynamic limit
($V \rightarrow \infty$ at constant fugacity or density)
the grand partition function factorizes
and yields a series of loop grand potentials.

Although that result appears both useful and formally exact,
in the present work a related but arguably more practical approach
is developed that is well-suited for computer simulation.
Since the non-zero contributions come only from compact loops,
one can restrict each of the sums above to configurations
where the consecutive particles around a loop are
actual spatial (or more generally phase space) neighbors.
This can be defined by some arbitrary cut-off
whose quantitative effect can be ascertained \emph{a posteriori}.
That is, in any configuration the symmetrization function is unity
plus the contributions from, and only from, loops
that are compact by  the imposed criterion.

\subsubsection{One-Dimensional Example}

In order to illustrate the idea concretely,
consider the case of four particles in one-dimension.
For four particles the symmetrization function in full is
\begin{eqnarray}
\lefteqn{
\eta^\pm({\bf \Gamma})
}  \\
& = &
1
\pm \left[ \eta_{q;12}^{(2)} + \eta_{q;23}^{(2)} + \eta_{q;34}^{(2)} \right]
+ \eta_{q;12}^{(2)}\eta_{q;34}^{(2)}
\nonumber \\ & & \mbox{ }
+ \left[ \eta_{q;123}^{(3)} + \eta_{q;234}^{(3)}  \right]
\pm \left[ \eta_{q;13}^{(2)} + \eta_{q;24}^{(2)} \right]
\nonumber \\ & & \mbox{ }
\pm \eta_{q;14}^{(2)}
+ \left[ \eta_{q;134}^{(3)} + \eta_{q;132}^{(3)}
+ \eta_{q;124}^{(3)} + \eta_{q;142}^{(3)}  \right]
\nonumber \\ & & \mbox{ }
\pm \eta_{q;1234}^{(4)}
+ \eta_{q;13}^{(2)}\eta_{q;24}^{(2)}
+ \eta_{q;14}^{(2)}\eta_{q;23}^{(2)}
+ \left[ \eta_{q;143}^{(3)} + \eta_{q;243}^{(3)}  \right]
\nonumber \\ & & \mbox{ }
\pm \left[ \eta_{q;1342}^{(4)}
+ \eta_{q;1423}^{(4)}
+ \eta_{q;1243}^{(4)} + \eta_{q;1432}^{(4)} + \eta_{q;1324}^{(4)}
 \right] .\nonumber
\end{eqnarray}
There is one monomer.
There are $4\times 3 /2 = 6$ single dimers.
There are 3 double dimers.
There are $4\times 3 \times 2 /3 = 8$ trimers.
There are $4\times 3 \times 2 \times 1 /4 = 6$ tetramers.
This gives $4!=24$ terms altogether.

In one dimension the symmetrization function
is dominated by nearest neighbor permutations.
Hence the terms are arranged here roughly in order of dominance.
As mentioned above,
this idea that the symmetrization function is dominated by
transpositions of neighbors carries over to two or more dimensions.

In simulations reported below
only the three groups corresponding
to the first line will be included.
For $N$ particles
define the monomer term,
\begin{equation}
\eta_{q;1}({\bf \Gamma}) \equiv 1 ,
\end{equation}
the nearest neighbor dimer,
\begin{eqnarray} \label{Eq:dim}
\eta_{q;2}({\bf \Gamma})
& = &
\sum_{j=1}^{N-1}
\eta_{q;j,j+1}^{(2)}
 \\ & = &
\sum_{j=1}^{N-1}
e^{-q_{j,j+1} p_{j,j+1} /i\hbar}  ,\nonumber
\end{eqnarray}
and the double nearest neighbor dimer,
\begin{eqnarray} \label{Eq:ddim}
\eta_{q;3}({\bf \Gamma})
& = &
\sum_{j=1}^{N-3}
\sum_{k=j+2}^{N-1}
\eta_{q;j,j+1}^{(2)}\eta_{q;k,k+1}^{(2)}
 \\ & = &
\sum_{j=1}^{N-3}
\sum_{k=j+2}^{N-1}
e^{-q_{j,j+1} p_{j,j+1} /i\hbar}
e^{-q_{k,k+1} p_{k,k+1} /i\hbar}  .\nonumber
\end{eqnarray}
Accordingly,
here  the $a$th order approximation to the symmetrization function
is defined as
\begin{equation}
\eta_{q;[a]}^\pm({\bf \Gamma})
=
\sum_{b=1}^a (\pm 1)^{b-1} \eta_{q;b}({\bf \Gamma}),
\;\; a=1,2,3.
\end{equation}
The case $a=1$ is for monomers,
also known as distinguishable particles.
This is the classical case in the event that the commutation function
is set to unity.
In some of the results below
including the single nearest neighbor dimer, $a=2$,
makes a measurable difference.
In almost no case was the additional contribution
from the double nearest neighbor dimer, $a=3$, measurable.

As mentioned above,
including these symmetrization functions does not preclude
the momentum integrals from being performed analytically.

%
\section{Computational Details}
\setcounter{equation}{0} \setcounter{subsubsection}{0}
%

\subsection{Model}

Results from the present mean field simulation approach
are tested against benchmark results given by
Hernando and Van\'i\v cek.\cite{Hernando13}
These are based on the first 50 energy eigenvalues,
which, apart from the  discretization of time and space,
are numerically exact.\cite{Hernando13}
The model used by these authors
consists of spinless Lennard-Jones particles
in one-dimension
trapped by a harmonic potential.
The latter is the harmonic oscillator potential
\begin{equation}
U_1({\bf q}) =
 \frac{1}{2} m \omega^2 \sum_{j=1}^N q_j^2 .
\end{equation}
The Lennard-Jones pair potential is
\begin{equation}
U_2({\bf q}) =
\epsilon \sum_{j<k}^N
\left[  \left(\frac{r_\mathrm{e}}{q_{jk}} \right)^{12}
 - 2 \left(\frac{r_\mathrm{e}}{q_{jk}} \right)^{6}  \right] ,
\end{equation}
where $q_{jk} =  q_{j} - q_{k}$.
Note that all particle pairs interact,
not just nearest neighbors.
The equilibrium separation $r_\mathrm{e}$ is related to the more usual
Lennard-Jones diameter $\sigma$ by $r_\mathrm{e} = 2^{1/6} \sigma$.

Hernando and Van\'i\v cek\cite{Hernando13} give results
at the de Boer quantum delocalization length
$L_\mathrm{dB} \equiv
2^{1/6}\hbar/(r_\mathrm{e}\sqrt{m\epsilon}) = 0.16$.
They state that this corresponds to hypothetical particles
between parahydrogen and neon.
They also use
$\omega r_\mathrm{e} \sqrt{m/\epsilon}=1/2$.
Unless stated otherwise,
parameters corresponding to these two values are used below.

In SI units, for this case I used
the neon mass, $m=3.351\times 10^{-26}$\,kg,
and Lennard-Jones diameter,
$r_\mathrm{e} = 2^{1/6} \sigma_\mathrm{Ne} = 3.13055\times 10^{-10}$\,m.
These give
$ \omega =  2^{1/6} \hbar^2 /(2 L_\mathrm{dB} r_\mathrm{e}^2 \hbar m )
= 1.1264 \times 10^{11}$\,s$^{-1}$,
and $\epsilon = 14.03 \, \hbar \omega = 1.6666\times 10^{-22}$\,J.
This last value is about one third of
the Lennard-Jones well-depth parameter for neon,
$\epsilon_\mathrm{Ne} = 4.93 \times 10^{-22}$\,J.
\cite{Sciver12}

The temperature below is reported
in terms of the frequency corresponding to the Lennard-Jones pair potential
at the equilibrium spacing,
$\omega_\mathrm{LJ} \equiv \sqrt{ u''(r_\mathrm{e})/m}
= \sqrt{72 \epsilon /m r_\mathrm{e}^2}
= 16.97 \omega$,
(since in the present simulations $\omega r_\mathrm{e} \sqrt{m/\epsilon}=1/2$).

\subsection{Simulation Algorithm}

The Metropolis Monte Carlo algorithm was used
in classical phase space
with umbrella sampling.\cite{Allen87}
The statistical average in the form of
the final phase space integral in Eq.~(\ref{Eq:<A>})
was written
\begin{equation}  \label{Eq:<AWh>/<Wh>}
\left<  A \right>_{T;W\eta^\pm}
=
\frac{\left<  A W\eta^\pm \right>_{T}}{\left< W\eta^\pm  \right>_{T}} .
\end{equation}
The right hand side 
consists of the ratio of two canonical averages in classical phase space,
which are well-suited to the Metropolis Monte Carlo algorithm.

For the product of single particle  commutation functions,
Eq.~(\ref{Eq:WSHO}),
with or without the symmetrization function,
Eqs~(\ref{Eq:dim}) or (\ref{Eq:ddim}),
and with the kinetic energy part of the Maxwell-Boltzmann factor,
it is evident that the momentum integrals
may be evaluated analytically.
The integrals are of the type
\begin{eqnarray}
\lefteqn{
\int_{-\infty}^\infty \mathrm{d} P \;
e^{-\alpha P^2/2} e^{iP  Q} P^n
} \nonumber \\
& = &
 \sqrt{2\pi} \alpha^{-(n+1)/2} e^{-  Q^2/2\alpha}
\nonumber \\ & & \mbox{ } \times
\sum_{m=0}^{\lfloor n/2 \rfloor}
\frac{n! (i Q/\surd \alpha)^{n-2m}}{(2m)!(n-2m)!} 
(2m-1)!! .
\end{eqnarray}
(Note here only, $(-1)!! \equiv 1$.)
Comparison between Monte Carlo in the full phase space,
and Monte Carlo in position space with the analytic quadratures,
showed agreement between the two,
with the latter method being the most reliable and
by far the most computationally efficient.
The results reported below
use the analytic momentum quadrature.

For the  single particle  commutation function,
Eq.~(\ref{Eq:WSHO}),
most results reported below used up to the sixth Hermite polynomial,
although tests with up to 12 terms were carried out.
Generally 6 iterates of Newton's method was used to locate each local minimum,
which is probably a factor of 2 too many.
No effort was made to optimize these parameters
or to adjust them for different temperatures or models.
A potential cut-off, generally $u_\mathrm{cut} = 5 \hbar \omega$,
was used so that $W_j({\bf p},{\bf q})  = 1$
whenever $U_j({\bf q}) - \overline  U_j({\bf q}) \ge U_\mathrm{cut}$.
This was also done if $\overline U''_j \le 0$.
The results were not sensitive to the value of the cut-off,
except that at the highest temperatures
large displacements $Q_j$ made a smaller value necessary,
$u_\mathrm{cut} \approx 2 \hbar \omega$.
In retrospect it might possibly have been
better to apply the cut-off criterion directly to the displacement $Q_j$.

High temperature expansions for the commutation function
were also implemented.\cite{QSM,STD2,Attard18a,Attard18b}
These involved up to the fourth tensor gradient of the potential.
The results were a little disappointing
in that their regime of validity,
$\beta \hbar \omega_\mathrm{LJ} \alt 1$,
is not easily resolved in the figures below.
These results are not reported here.

The implementation of the Metropolis algorithm
for a classical canonical system was standard
and need not be repeated here.
The simulation was broken into 500 blocks,
and the statistical error was estimated
from the fluctuations in the ratio of the averages of each block.
The error reported below is twice the standard deviation
divided by the square root of the number of blocks,
which corresponds to the 96\% confidence level.
This calculated error
was consistent with the error calculated from the fluctuations
between independent repeat runs.

The computations were performed on a teen-aged  personal computer.
Typically a given simulation for five particles
took between two minutes and two hours,
depending on the temperature and the accuracy sought.
Turning off the calculation of the symmetrization function
reduced the computation time by about a factor of two
in one typical case.

Some relatively short tests were carried out for the system size dependence.
With $4 \Rightarrow 8 \Rightarrow 16 \Rightarrow 32 \Rightarrow 64$ particles,
the computer time increased consecutively
by factors of 1.9, 2.2, 2.7, and 3.9 respectively.
The case of $N=64$ took 303 seconds
for a relative statistical error in the total energy of 0.03\%.
In these five cases only nearest and next nearest neighbors
were included in the Lennard-Jones potential calculation,
which corresponds to a cut-off of
$R_\mathrm{cut} \approx $2.5--3$r_\mathrm{e}$.
(No cut-off was used in any other results reported in this paper.)
These particular tests used the same number of trial position cycles
(a cycle is one attempted move of every particle),
and hence the number of trial positions increased linearly
with particle number.
With the cut-off and effective neighbor table,
the cost of a trial move of one particle
is independent of system size.
In the simulations, after every six cycles of trial moves,
the commutation function and the symmetrization function were calculated,
and the quantities to be averaged were accumulated.
The costs of evaluating
a single particle commutation function
and a nearest neighbor dimer symmetrization function,
are independent of the size of the system.

There are $\sim N$ nearest neighbor dimers,
and $\sim N^2/2$ double nearest neighbor dimers.
Therefore the cost of evaluating the instantaneous weighted potential energy
at these two levels scales as $N$ and $N^2$, respectively.
The cost of evaluating the  instantaneous weighted kinetic energy
scales as $N^2$ and $N^3$, respectively,
because of the momentum integrals.
These are the dominant costs of the simulation;
the cost of a run is the number of cycles times these.

This assumes an inhomogeneous system in which each particle
makes a distinct contribution that has to be explicitly accounted for,
which is what is done here.
In principle, for a homogeneous system, the kinetic energy
could be obtained by calculating
one representative term of each type that can occur,
and simply multiplying it by an appropriate combinatorial factor,
which would mean that
the cost of evaluating the  instantaneous weighted kinetic energy
would not depend on $N$.
Of course this would mean that one would have to increase
the number of cycles of trial configurations
in order to obtain  statistical accuracy
comparable to the present case when each contribution is evaluated explicitly
and accumulated.

The relative statistical error in the total energy
in the present specific tests at constant number of trial position cycles
was
8.7\%, 2.7\%, 0.20\%, 0.11\%, and 0.03\%, respectively.
In general the statistical error scales inversely with the square
root of the computer time, all other things being equal.
These values mean that one would have to run the present $N=4$ system
2,700 times as long a time as the present $N=64$ system was run
in order to get the same statistical error.

\comment{ 
The present algorithm suffers from a computer limitation
for larger number of particles.
The single particle mean field commutation functions
can be less than unity,
and so for large $N$ the full function,
which is their product $W({\bf \Gamma}) =\prod_{j=1}^N W_j({\bf \Gamma})$,
can be zero within the computer's range of digital numbers.
There appear to be three possible hacks.
One could rescale the single particle mean field commutation functions
by a fixed constant.
Or else one could incorporate the magnitude of the commutation function
into the Metropolis algorithm.
Or else one could take only the product of those single particle
commutation functions that are directly involved
in the specific term of each function being averaged.
The first of these fixes was successfully tested.
} 

%
\section{Results}
\setcounter{equation}{0} \setcounter{subsubsection}{0}
%

\subsubsection{Neon--Parahydrogen}

\begin{figure}[t!]
\centerline{
\resizebox{8cm}{!}{ \includegraphics*{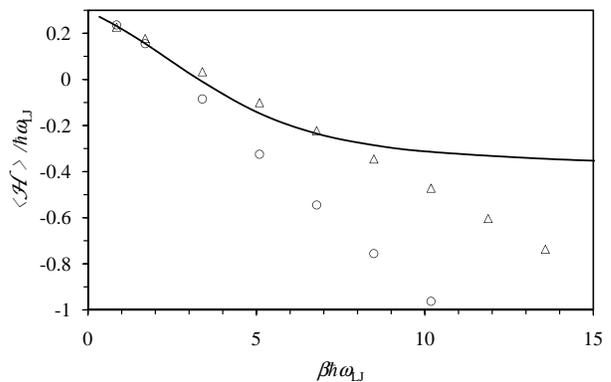} } }
\caption{\label{Fig:EvB4}
Average energy as a function of inverse temperature for
$N=4$ distinguishable spinless Lennard-Jones neon--parahydrogen particles
in a harmonic trap in one-dimension.
The solid curve is the exact result  for particles
derived from the first 50 energy eigenvalues
obtained by Hernando and Van\'i\v cek,\cite{Hernando13}
the circles are the classical result,
and the triangles are the quantum result
in mean field simple harmonic oscillator monomer approximation
(ie.\ without symmetrization).
The statistical error is less than the symbol size.
}
\end{figure}

Results for the average energy of $N=4$ Lennard-Jones particles
are shown in Fig.~\ref{Fig:EvB4}.
These are for
$L_\mathrm{dB}  = 0.16$ and
$\omega r_\mathrm{e} \sqrt{m/\epsilon}=1/2$,
which correspond to neon--parahydrogen.\cite{Hernando13}
The Lennard-Jones frequency used as a scale is much larger
than that of the harmonic potential, $\omega_\mathrm{LJ} = 16.97 \omega$.
The exact results were derived by computing the canonical average
of the first 50 energy eigenvalues
obtained by Hernando and Van\'i\v cek.\cite{Hernando13}
Since originally an unknown constant has  been applied
to shift the eigenvalues 
to the Lennard-Jones potential minimum,\cite{Hernando13}
here the exact average energy curve has been shifted
into coincidence with the classical result
at the highest temperature studied.

It can be seen that the classical results remain accurate
for temperatures  $\beta \hbar \omega_\mathrm{LJ} \alt 2$.
The high temperature expansion (not shown)
is close to the classical result.
The mean field quantum theory has a larger regime of validity,
and it remains accurate
for temperatures  $\beta \hbar \omega_\mathrm{LJ} \alt 7$.

Interestingly enough,
the classical equipartition theorem
for the kinetic energy is not satisfied in a quantum system.
It increasingly underestimates the actual kinetic energy
as the temperature is lowered.
For example, at $\beta \hbar \omega_\mathrm{LJ} = 17$,
for $N=4$ the mean field theory for indistinguishable particles
gives the kinetic energy as $\beta {\cal K}/N = 1.83 \pm 0.03$.

In the present model symmetrization effects were practically negligible.
For example, at $\beta \hbar \omega_\mathrm{LJ} = 17$,
adding the single nearest neighbor dimer to the symmetrization function
decreased the average energy for bosons by about 0.2\%,
and increased that for fermions by the same amount.
Adding to this the double nearest neighbor dimer contribution
increased the energy by about 6 parts in $10^6$.
At this temperature, the thermal wave length is only slightly
larger than the equilibrium spacing,
$\Lambda_\mathrm{th} = 1.34 r_\mathrm{e}$,
and so one expects symmetrization effects to be small.

(The momentum integrals for a dimer applied to the classical case
$W=1$ yield Gaussians
$\exp - 2\pi q_{jk}^2/\Lambda_\mathrm{th}^2$,
which is why symmetrization effects are negligible
unless the typical particle spacing is somewhat
smaller than the thermal wavelength.)

\begin{figure}[t!]
\centerline{
\resizebox{8cm}{!}{ \includegraphics*{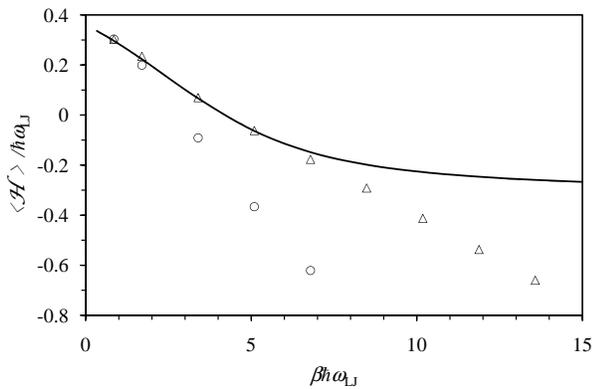} } }
\caption{\label{Fig:EvB5}
Same as preceding figure but for $N=5$.
}
\end{figure}

The average energy for $N=5$ is shown in Fig.~\ref{Fig:EvB5}.
The relative performance of the classical and mean field quantum results
is similar to that in the $N=4$ case,
and their regime of accuracy is about the same.
Symmetrization effects were also quantitatively similar
to that of the previous case.
This is perhaps not surprising since the thermal wave length is unchanged.

\begin{figure}[t!]
\centerline{
\resizebox{8cm}{!}{ \includegraphics*{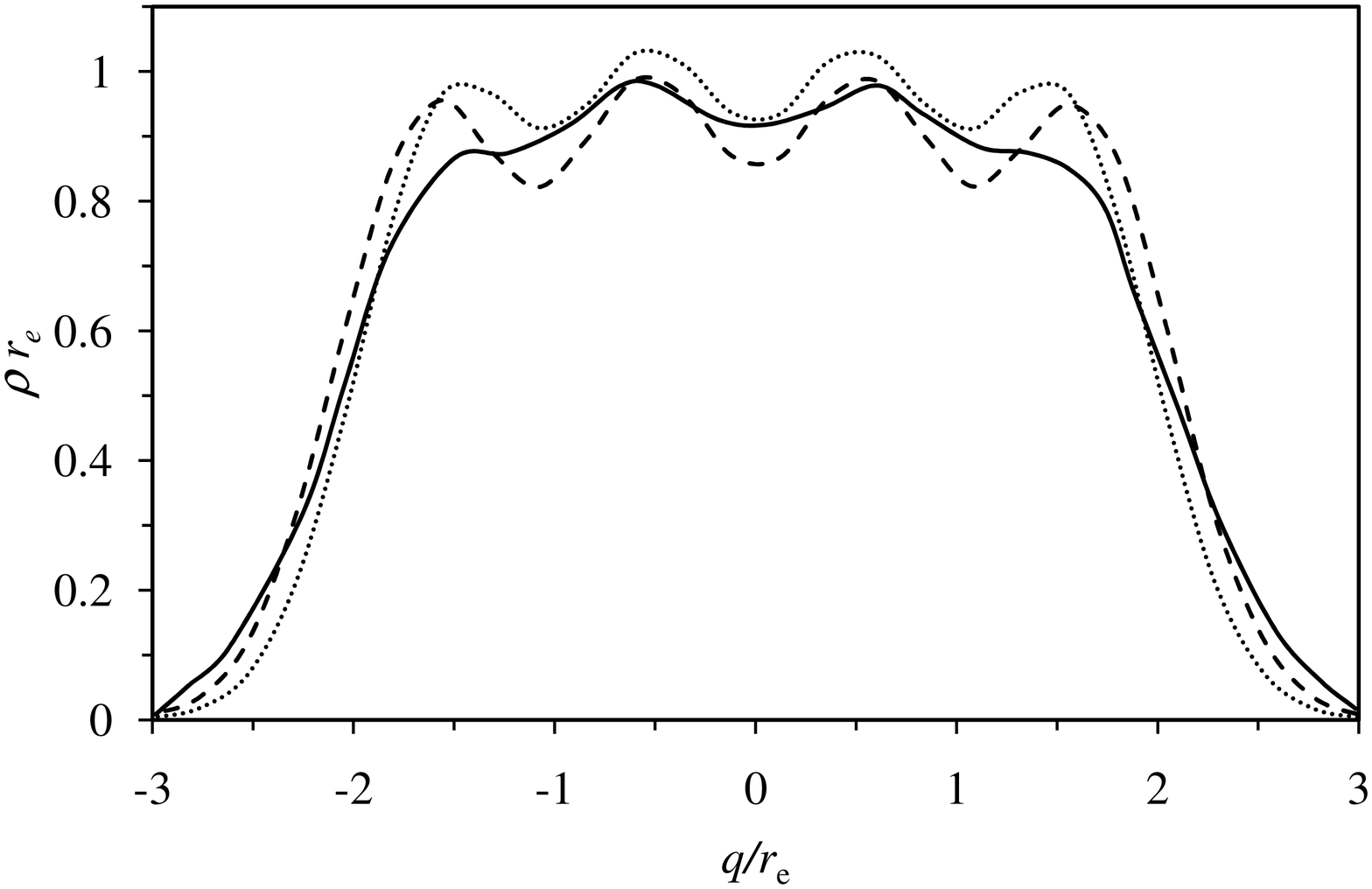} } }
\caption{\label{Fig:rhoq.4}
Density profile for $N=4$ and $\beta \hbar \omega_\mathrm{LJ} = 6.8$.
The solid curve is the exact result for distinguishable particles,
\cite{Hernando13}
the dotted curve is the classical result,
and the dashed curve is the  mean field monomer quantum result.
}
\end{figure}

Figure~\ref{Fig:rhoq.4} compares the density profiles
for $N=4$  and $\beta \hbar \omega_\mathrm{LJ}  = 6.8$.
The neon--parahydrogen particles are indistinguishable;
symmetrization effects on the density profile
are almost immeasurable in this case.
The exact result\cite{Hernando13}
for distinguishable particles shows two relatively broad central peaks
and two outer shoulders.
The peaks in the  present mean field monomer quantum calculations
are more pronounced than in the exact calculations,
as are those in the classical result.
It is noticeable how much of the density profile
for the neon--parahydrogen particles
is purely classical.
The quantum density profile spreads out
more than the classical profile,
and the average density in the region
$-2r_\mathrm{e} \alt q \alt 2r_\mathrm{e}$ is lower.
This indicates that the harmonic trap confines the classical particles
to a smaller region than it does the quantum particles.

\begin{figure}[t!]
\centerline{
\resizebox{8cm}{!}{ \includegraphics*{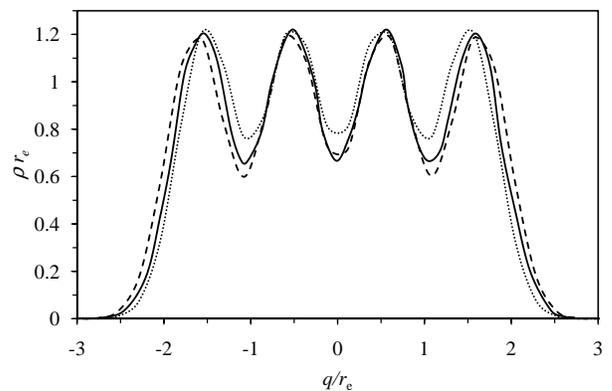} } }
\caption{\label{Fig:rhoq.7}
Same as preceding figure but at
$\beta \hbar \omega_\mathrm{LJ} =11.9$.  
}
\end{figure}

Figure~\ref{Fig:rhoq.7}
is for the lower temperature of $\beta \hbar \omega_\mathrm{LJ}  = 11.9$.
The peaks in the density profile are now quite pronounced
in  all three approaches.
Again one sees that the major contribution to the density profile
is classical.
The mean field approximation gives quantitatively
the quantum deepening of the density minima.
At this temperature
the mean field theory is in better agreement
with the exact results\cite{Hernando13}
than at the higher temperature of the preceding figure.
Indeed, comparing these results to those in Fig.~\ref{Fig:EvB4},
one can conclude that
the regime of accuracy for the mean field approximation appears to be
greater for structure than it is for energy.

\begin{figure}[t!]
\centerline{
\resizebox{8cm}{!}{ \includegraphics*{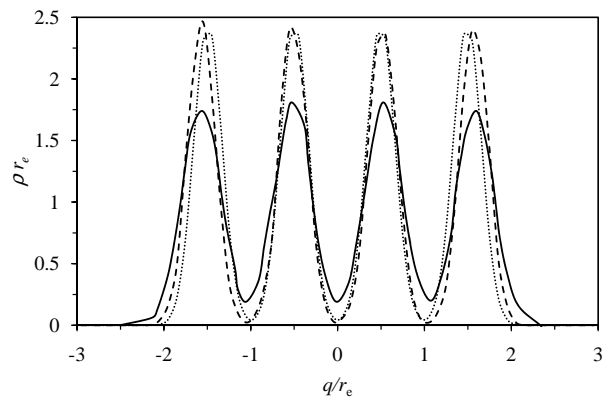} } }
\caption{\label{Fig:rho2.8}
Same as preceding figure but at
$\beta \hbar \omega_\mathrm{LJ}  = 47.5$. 
}
\end{figure}

Figure~\ref{Fig:rho2.8}
is the density profile
for the still lower temperature of $\beta \hbar \omega_\mathrm{LJ} = 47.5$.
In this case the mean field quantum result is almost coincident
with the classical result.
Both magnify the structure that is apparent in the exact results.
\cite{Hernando13}
Including symmetrization effects would make no apparent difference
to the mean field quantum results.

The peaks in the density profile are more pronounced
at this low temperature than at the higher temperatures
of the preceding figures.
The spacing of the peaks is about $r_\mathrm{e}$
in all three cases.
In this system confined by the harmonic potential,
the peak spacing is relatively insensitive to the number of particles
since extra particles just spread further in the harmonic potential trap.
Even for a relatively high number of particles
the average nearest neighbor spacing
is not much less than $r_\mathrm{e}$.
(For 64 helium particles at $\beta \hbar \omega_\mathrm{LJ} = 17$,
the spacing at the ends is about $r_\mathrm{e}$,
and the spacing in the center is about $0.8 r_\mathrm{e}$.)

\subsubsection{Helium}

Results were also obtained for helium-4,
using $m = 6.65 \times 10^{-27}$\,kg,
and Lennard-Jones parameters
$\epsilon_\mathrm{He} =   1.41\times 10^{-22}$\,J,
and $\sigma_\mathrm{He}  = 2.56\times 10^{-10}$\,m.\cite{Sciver12}
These correspond to a de Boer length of
$L_\mathrm{dB} = 0.426$.
The dimensionless trap frequency was unchanged,
$\omega r_\mathrm{e} \sqrt{m/\epsilon}=1/2$,
which corresponds to  $\omega = 2.54\times 10^{11}$\,Hz.

\begin{figure}[t!]
\centerline{
\resizebox{8cm}{!}{ \includegraphics*{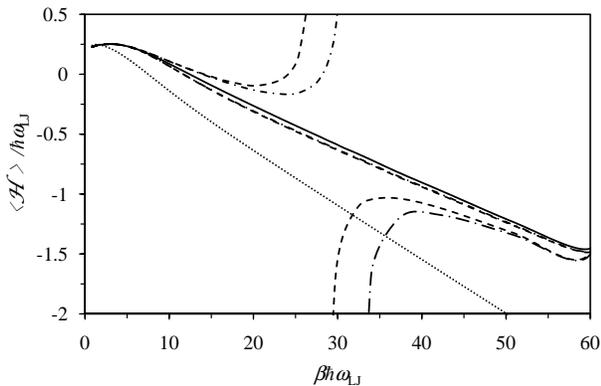} } }
\caption{\label{Fig:HvB-He}
Average energy for helium, $N=4$.
The dotted curve is  the classical result,
the solid curve is the mean field quantum result for monomers,
the dashed curves include nearest neighbor dimers,
and the dash-dotted curves include double nearest neighbor dimers.
At high temperatures,
the upper pair of curves is for fermions,
and the lower (coincident) pair is for bosons.
The statistical error is about the width of the curves
everywhere except in the vicinity of the pole.
}
\end{figure}

Figure~\ref{Fig:HvB-He}
shows the average total energy at a function of inverse temperature
in the classical,
quantum mean field monomer,
nearest neighbor dimer,
and double nearest neighbor  dimer cases.
Unlike the case of neon--parahydrogen treated above,
for helium there is a measurable difference
between distinguishable particles,
bosons, and fermions.
(Of course $^4$He is a boson;
the fermion results here are used to illustrate the effects of particle
statistics rather than as a quantitative application
to a real physical system.)

As for neon--parahydrogen,
quantum effects in helium increase the energy over its classical value.
Further, adding wave function symmetrization
decreases the energy for bosons
and generally increases the energy for fermions.
This effect was not noticeable  for Ne--H$_2$,
Fig.~\ref{Fig:EvB4}.
At  $\beta \hbar \omega_\mathrm{LJ}  = 17$,
the thermal wave-length is
$2.2r_\mathrm{e}$ for He,
and $1.3r_\mathrm{e}$ for  Ne--H$_2$.

At around $\beta \hbar \omega_\mathrm{LJ}  = $ 28--29 (nearest neighbor dimer),
there is a singularity in the results for fermions.
This is due to the denominator in the umbrella average,
Eq.~(\ref{Eq:<AWh>/<Wh>}), passing through zero
when nearest neighbor dimers are used
(ie.\ $\eta^-_{q;[2]}$, the upper dashed curve in Fig.~\ref{Fig:HvB-He}).
Adding  double nearest neighbor  dimers,
(ie.\ $\eta^-_{q;[3]}$, the upper dash-dotted curve)
shifts the singularity to lower temperatures, but does not remove it.
Of course in general whenever the dominant term
in a convergent expansion vanishes,
one has to go to otherwise negligible higher order terms
to get reliable results.
One might have to include further terms in the symmetrization expansion
to delineate the behavior here more reliably.

Just beyond the range of temperatures plotted
the quantum energy curves become structured.
This seems to arise from the commutation function,
and is most likely just a low temperature artifact of the mean field
simple harmonic oscillator approximation.

\begin{figure}[t!]
\centerline{
\resizebox{8cm}{!}{ \includegraphics*{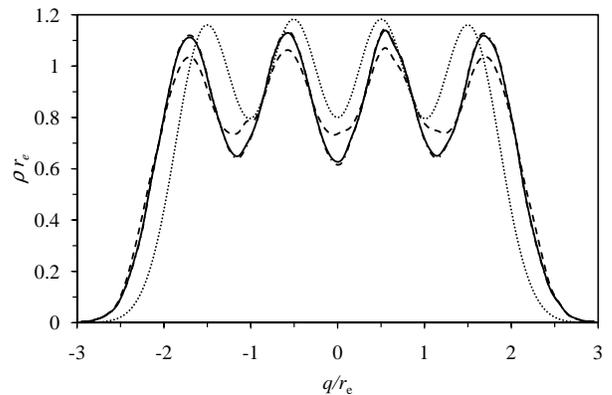} } }
\caption{\label{Fig:rho-He}
Density profile for helium at $N=4$ and
$\beta \hbar \omega_\mathrm{LJ}  = 27$.
The dotted curve is the classical result,
the solid curve is the mean field quantum result for monomers,
the dashed (fermions) and dash-dotted (bosons) curves
both include nearest neighbor and double nearest neighbor dimers.
The results for bosons are virtually indistinguishable
from those for distinguishable particles.
}
\end{figure}

Figure~\ref{Fig:rho-He} shows the density profile
at  $\beta \hbar \omega_\mathrm{LJ} =27$, 
near the fermion singularity.
As in the earlier density profiles for  Ne--H$_2$,
classical effects dominate the structure of the system,
even for He.
Also like the earlier results,
quantum effects spread the overall density profile
so that the quantum particles tend to be less confined
by the harmonic potential trap than their classical counterparts.

The thermal wavelength here is $\Lambda_\mathrm{th}=2.8 r_\mathrm{e}$,
whereas the spacing between density peaks is almost exactly $r_\mathrm{e}$.
As mentioned above,
an estimate of the magnitude of symmetrization effects
is provided by the Gaussian
$\exp - 2\pi q_{j,j+1}^2/\Lambda_\mathrm{th}^2
\approx \exp - 2\pi /2.8^2 = 0.45$.
One might therefore expect some symmetrization effects here.
The fact that  in Fig.~\ref{Fig:rho-He}
the profile for bosons is virtually unchanged
from that of distinguishable particles
suggests that the induced pair attraction due to being able
to occupy the same state is weak compared to the short-ranged repulsion
of the pair potential.
Boson symmetrization has a more noticeable effect on the energy,
Fig.~\ref{Fig:HvB-He}.

That the results for indistinguishable fermions are distinguishable here
is due to the fact that the temperature is close to the fermionic pole
that was discussed in conjunction with Fig.~\ref{Fig:HvB-He}.
The nature of fermion statistics is to reduce the height of the peaks
in the density profile
and to reduce correspondingly the depth of the valleys.
This can be understood from the induced repulsion
due to fermions not being able to occupy the same state.
This has the effect that two fermions tend not to simultaneously
occupy positions corresponding to adjacent peaks,
which would otherwise be their most likely locations.
Hence when one fermion is at a peak,
the neighboring ones are more likely to be pushed toward the further valley
compared with the situation for bosons.
The  effects of particle statistics
on the present density profile of He fermions
at  $\beta \hbar \omega_\mathrm{LJ} =27$ 
is qualitatively similar to that for Ne--H$_2$ fermions
at zero temperature given by
Hernando and Van\'i\v cek in their Fig.~1.\cite{Hernando13}

The quantum density profiles in  Fig.~\ref{Fig:rho-He}
are more noisy than their classical counterpart.
This is probably due to the umbrella sampling.
In principle it is possible to avoid umbrella sampling
by, after the momentum integration,
incorporating the magnitudes of the commutation function
and the symmetrization function into the Maxwell-Boltzmann phase space weight,
and factoring their sign into the function being averaged.
(The Metropolis and other algorithms require a positive weight density.)

%
\section{Conclusion}
\renewcommand{\theequation}{\arabic{section}.\arabic{equation}}
%

This paper has tested the author's mean field quantum harmonic oscillator
algorithm against the benchmark results of
Hernando and Van\'i\v cek \cite{Hernando13}
for one-dimensional interacting Lennard-Jones
neon-parahydrogen particles.
It was found that the mean field approximation was quantitatively accurate
down to the relatively low temperatures
of
$\beta \hbar \omega_\mathrm{LJ}  \alt 7$ for the average energy,
and
$\beta \hbar \omega_\mathrm{LJ}  \alt 12$ for the average density profile.
In contrast the classical theory
and the high temperature quantum expansion
were only valid for the average energy
for $\beta \hbar \omega_\mathrm{LJ}  \alt 2$.

It was straightforward to program the method
as a modification of existing Monte Carlo simulation algorithms
in classical phase space.
The computational requirements of the present approach
were not high,
with most runs being completed in minutes on a personal computer.

It is likely that the present
one dimensional  tests of the mean field theory are overly pessimistic.
Since the fluctuations that are neglected by the mean field
decrease with the square root of the number of interacting particles,
the theory becomes more accurate
as the range and magnitude of the potential is increased,
the temperature is decreased,
the density is increased,
and the dimensionality is increased.
Hence one would expect the present theory
to perform better in three dimensions
than the present one-dimensional results indicate.

The second approximation
in the present algorithm was to
map each configuration to that of independent harmonic oscillators
based on the instantaneous local potential minimum
for each individual particle.
Modeling each particle as if it were instantaneously trapped
by the other particles is reasonable because of the high frequency
of the effective oscillator compared to the collective motions
of the particles.
The independent harmonic oscillator approximation
enabled the commutation function for the actual configuration
to be written as the product of the exact
simple harmonic oscillator commutation functions.
The present results indicate that
this approach has a larger range of utility
than the high temperature expansions.
\cite{Wigner32,Kirkwood33,QSM,STD2,Attard18b}

Finally the present results were in one dimension and
for a relatively small number of particles.
There is nothing intrinsic in the algorithm
that would restrict it to this regime.
In particular,
both the  commutation function
and the symmetrization function
are readily obtained in three dimensions.
With the use of a potential cut-off and neighbor tables,
both will likely have favorable scaling properties with system size.


\end{document}